# Pulsed coherent spectroscopy of a quantum emitter in hexagonal Boron Nitride

Jake Horder[1,2], Hugo Quard[1,2,†], Kenji Watanabe[3], Takashi Taniguchi[4], Nathan Coste[1,2], Igor Aharonovich[1,2,*]

[1] School of Mathematical and Physical Sciences, University of Technology Sydney, Ultimo, New South Wales 2007, Australia

[2] ARC center of Excellence for Transformative Meta-Optical Systems, University of Technology Sydney, Ultimo, New South Wales 2007, Australia

[3] Research Center for Functional Materials, National Institute for Materials Science, Tsukuba 305-0044, Japan

[4] International Center for Materials Nanoarchitectonics, National Institute for Materials Science, Tsukuba 305-0044, Japan

* Corresponding author: *igor.aharonovich@uts.edu.au, †hugo.quard@uts.edu.au

**Abstract**

*Defects in solid-state systems constitute a promising platform for the realization of deterministic quantum emitters. Among many candidate materials and emitters, point defects in hexagonal Boron Nitride (hBN) have recently emerged as particularly promising. In this work, we probe the coherence of an individual B center with a zero phonon line at 436 nm, under pulsed resonant excitation. We observe power-dependent Rabi oscillations up to $5\pi$, demonstrating optical coherent control of the transition. We achieve an excellent single photon purity of 93% at $\pi$-pulse. Furthermore, we probe the coherence of the two-level system using Ramsey interferometry, revealing an inhomogeneous coherence time of $T_2^* = 0.60$ ns. These results establish B centers in hBN as viable candidates for triggered, coherent quantum emitters and represent an important step towards their integration into quantum photonic platforms.*

**Manuscript**

Quantum emitters are a fundamental building block for quantum technologies, enabling applications in quantum communication, photonic quantum computing, and quantum metrology[1,2]. Solid-state quantum emitters offer a particularly attractive route towards scalable quantum emitters[3,4], combining deterministic photon generation with the potential for on-chip integration. Several material platforms have been investigated, including semiconductor quantum dots[5–7], single molecules[8–10] and color centers in wide-bandgap materials[11–14]. However, the majority of these systems still face significant challenges, such as spectral inhomogeneity across emitters, limited control over defect creation, and constraints in material processing and integration. These limitations hinder large-scale scalability and motivate the search for alternative solid-state platforms with improved uniformity and manufacturability.

Hexagonal boron nitride (hBN) has recently emerged as a versatile nanophotonic platform. Its layered, van der Waals structure is particularly appealing for the realization of photonic nanostructures and integration within quantum nanophotonic platforms. Further, it is an excellent material for quantum emitters operating from the visible to the near-infrared[15,16]. Particularly, the B center, with a zero phonon line (ZPL) at 436 nm, has attracted considerable interest due to its relatively narrow and reproducible optical linewidth[17–19], precise creation via electron-beam irradiation[20,21], and compatibility with nanophotonic integration[22,23]. Indeed, recently two-photon interference from the B center has been realized[24], as well as the direct observation of its defect structure[25].

Here we explore the operation of a single B center in the regime of ultrafast pulsed resonant excitation. We demonstrate coherent population control through power-dependent Rabi oscillations with pulse areas up to 5π, confirming the two-level nature of the optical transition. At π-pulse, we measure a high single-photon purity of 93%, confirming the single-photon nature of the emission. Furthermore, we employ Ramsey interferometry to directly probe the optical coherence of the emitter, extracting an inhomogeneous coherence time of $T_2^* = 0.60$ ns. These results establish B centers in hBN as coherent, triggered quantum emitters and highlight their potential for future quantum photonic applications.

Figure 1a depicts the physical system and excitation configuration used in this work. The sample consists of a high-purity hBN crystal grown by High-Pressure High-Temperature (HPHT), from which a thin flake was exfoliated onto a $SiO_2$/Si substrate. B centers were created by electron-beam irradiation and subsequently addressed optically either with an off-resonant 405 nm pump or with a resonant laser tuned to the 436 nm ZPL of the defect. Excitation of a single B center leads to single-photon emission either directly into the ZPL or into the phonon sideband (PSB) at lower photon energies due to coupling between the defect and the lattice vibrations. The ensemble photoluminescence spectra in Figure 1b clearly illustrate these processes. Under off-resonant excitation (blue curve), the emission is dominated by a narrow ZPL at 436 nm and accompanied by a broad PSB centered around 460 nm. Under resonant excitation (orange curve), scattered pump light spectrally overlaps with the ZPL. As a result, direct detection of the resonant ZPL emission is experimentally challenging, requiring high-extinction cross-polarization, whereas the PSB can be isolated by simple spectral filtering. This makes the PSB an experimentally robust and convenient detection channel for probing the defect's response in resonant measurements.

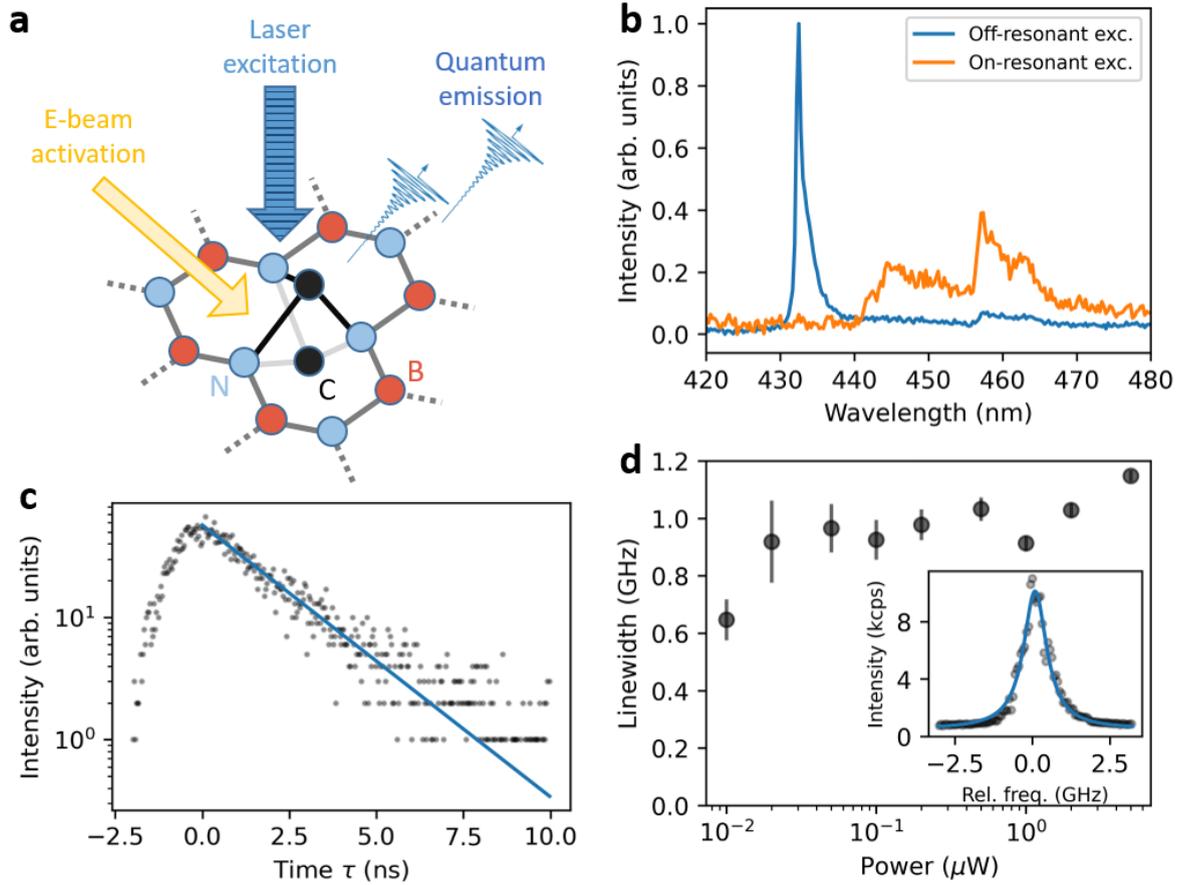

*Figure 1. Optical characterization of a single B center defect in hBN. a) Schematic of the carbon split-interstitial B center defect formed in hBN following electron-beam irradiation. The emitter is driven either off-resonantly or resonantly (cw or pulsed) and produces single-photon emission into its ZPL and PSB. b) Ensemble photoluminescence spectra of B centers. Off-resonant excitation at 405 nm (blue) reveals a sharp ZPL at 436 nm and a broad PSB near 460 nm. Under resonant excitation at 436 nm (orange), collection through a 440 nm long-pass filter suppresses scattered pump light, isolating only the PSB. c) Time-resolved photoluminescence from the investigated single B center under pulsed off-resonant excitation. An exponential fit yields a radiative lifetime of 1.95 ns. d) Power-dependent resonant response. The main panel shows the absorption linewidth as a function of excitation power. The inset displays a PLE spectrum recorded at the saturation power 4.41 µW, obtained by scanning a narrow-linewidth laser across the ZPL while collecting PSB emission, yielding an inhomogeneously broadened linewidth of 970 MHz.*

We then turn to the characterization of a single B center. Figure 1c shows the time-resolved photoluminescence trace measured under off-resonant excitation, from which we extract a radiative lifetime of $T_1 = 1.95$ ns. This lifetime corresponds to a Fourier-limited homogeneous linewidth of $\Gamma_n = 1/(2\pi T_1) \approx 82$ MHz, providing a reference scale for the intrinsic transition linewidth in the absence of dephasing. To probe the actual optical linewidth, we perform photoluminescence-excitation (PLE) spectroscopy: a narrow-linewidth laser is scanned across the ZPL while only the PSB emission is detected. The PLE response, shown in the inset of Figure 1d, directly maps the frequency-dependent absorption of the defect and reveals a linewidth of approximately 970 MHz for excitation powers around the saturation threshold of 4.41 µW. Increasing the excitation power leads to power broadening, but the linewidth remains close to 1 GHz throughout the accessible range, as shown in the main panel of Figure 1d. This contrast between the lifetime-limited linewidth and the measured PLE linewidth reflects the spectral diffusion characteristic of B centers in pristine hBN[26], with an order-of-magnitude broadening that nonetheless remains sufficiently small to enable the observation of B center electronic coherence.

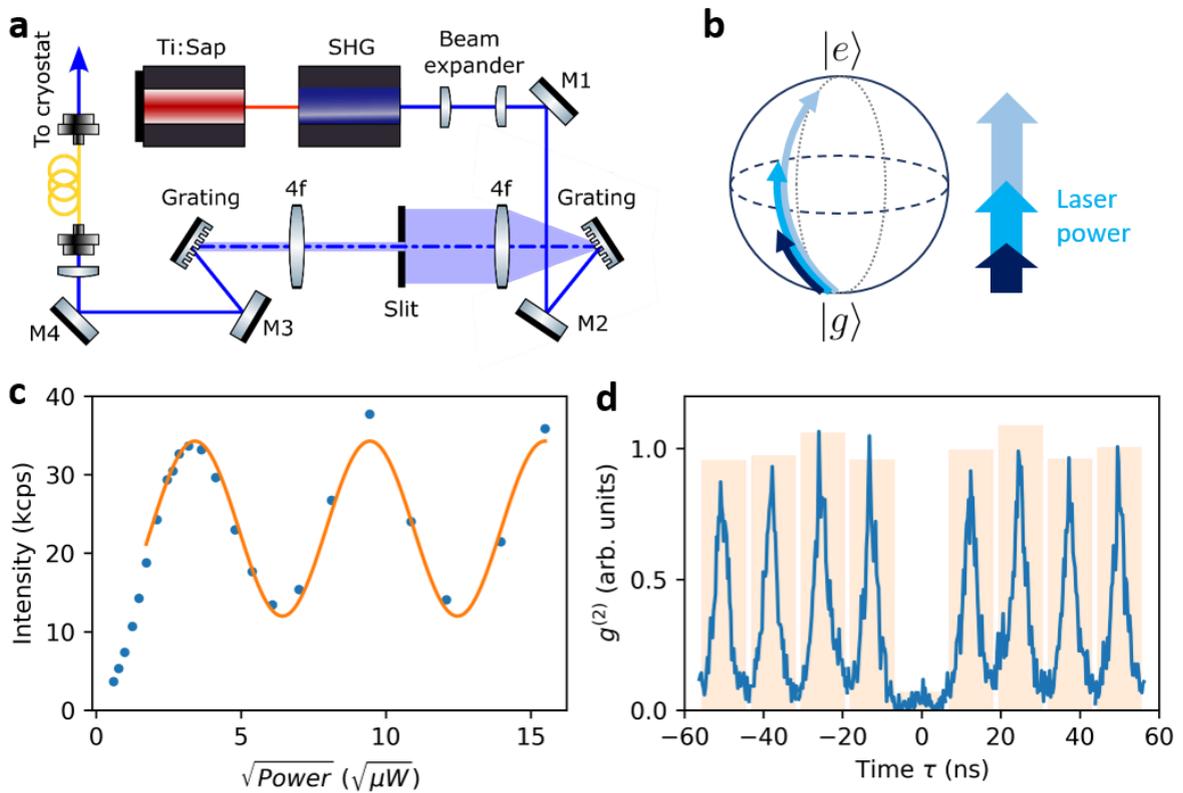

*Figure 2. Rabi control of a single B center emitter. a) Free-space optical setup implementing a 4f pulse-shaping stage to spectrally narrow the resonant excitation pulses. b) Rabi-driving scheme on the Bloch sphere, where increasing pulse power rotates the Bloch vector from the ground state |g⟩ toward the excited state |e⟩; a π-pulse corresponds to full inversion. c) Resonant PSB fluorescence as a function of the square root of the pulse power, showing clear Rabi oscillations up to 5π, with the first π-rotation occurring at 11.63 µW. d) Second-order autocorrelation measurement under π-pulse excitation. The strongly suppressed central peak yields $g^{(2)}(0)$=0.07 without background correction, confirming high-purity single-photon emission.*

To probe the superposition between the ground state |g⟩ and excited state |e⟩ of the B center under resonant excitation, we drive its optical transition using spectrally shaped pulsed light and detect the resulting PSB emission. A frequency-doubled femtosecond Ti:Sapphire laser operating at 80 MHz and centered at 436 nm provides the resonant pulses. Its native 2 nm

bandwidth is reduced to the spectrometer-limited 300 GHz (≈0.19 nm) using a 4f grating-based pulse-shaping stage with a slit placed at the Fourier plane of the first lens (Figure 2a). The spectrally narrowed pulses are then fiber-coupled and launched into the free-space cryogenic photoluminescence setup.

To drive Rabi oscillations, we keep the pulse duration fixed and vary the excitation power. The B center behaves as an effective two-level system with ground and excited states, $|g\rangle$ and $|e\rangle$, driven by a resonant electric field. In the rotating frame and under the rotating-wave approximation, the interaction Hamiltonian takes the standard form

$$\hat{H} = \frac{\hbar}{2}\left(\Omega\hat{\sigma}_x - \Delta\hat{\sigma}_z\right)$$

where $\Delta$ is the detuning between the transition frequency and the laser frequency, $\Omega = dE_0/\hbar$ is the resonant Rabi frequency set by the field amplitude $E_0$ and transition dipole moment $d$, and $\hat{\sigma}_i$ are the Pauli matrices. Solving the Schrodinger equation for a system initially in $|g\rangle$ and assuming resonant driving ($\Delta = 0$) yields the excited-state amplitude

$$c_e(t) = -i\sin\left(\frac{\Omega}{2}t\right)$$

Because the Rabi oscillation scales as $\Omega \propto E_0 \propto \sqrt{P}$, where $P$ is the optical power, the excited-state population after a pulse of fixed duration $\tau$ becomes

$$P_e(P) = |c_e(\tau)|^2 = \sin^2\left(\alpha\sqrt{P}\right)$$

with $\alpha$ a constant depending on the pulse duration. Increasing the excitation power therefore induces progressively larger rotations of the Bloch vector, promoting the emitter from $|g\rangle$ toward $|e\rangle$, as illustrated in Figure 2b. A π-pulse corresponds to a rotation that completely inverts the population from the ground state into the excited state. The full derivation is provided in the Supplementary Information.

Figure 2c shows the PSB fluorescence as a function of $\sqrt{P}$, directly reflecting this pulse-area scaling. Clear Rabi oscillations are observed up to approximately 5π with a power of 250 µW, with the first π-pulse obtained at 11.63 µW. The oscillation minima do not reach zero intensity, which we attribute to phonon-assisted excitation, and also to slow spectral wandering of the emitter during acquisition, which introduces slight detuning fluctuations. Both effects are common in solid-state defects and lift the minima of the Rabi oscillations. The single-photon purity under π-pulse excitation is evaluated from the second-order autocorrelation of the PSB emission (Figure 2d). We obtain a raw purity of 93%, without background correction, demonstrating that the B center generates high-quality single photons under resonant pulsed excitation.

Having calibrated the π-pulse power from the Rabi measurement, we then perform Ramsey interferometry to probe the coherence of the electronic superposition state. The sequence consists of two resonant π/2 pulses separated by a variable free-evolution delay τ. Starting from the ground state $|g\rangle$, the first $\pi/2$ pulse applies a rotation around the $x$-axis of the Bloch sphere

$$\hat{U}_{\pi/2} = \frac{1}{\sqrt{2}}\left(\hat{I} + i\hat{\sigma}_x\right)$$

preparing the emitter in the coherent superposition

$$|\psi\rangle_1 = \frac{1}{\sqrt{2}}(|g\rangle + i|e\rangle)$$

During the delay $\tau$, the superposition acquires a relative phase at the optical transition frequency $\omega_0$

$$|\psi\rangle_2 = \frac{1}{\sqrt{2}}\left(|g\rangle + ie^{i\omega_0\tau}|e\rangle\right)$$

corresponding to free precession of the Bloch vector around the equator, as depicted in Figure 3a. A second $\pi/2$ pulse maps this accumulated phase onto the excited-state population, yielding

$$P_e(\tau) = \frac{1}{2}\left(1 + \cos(\omega_0\tau)\right)$$

which produces the characteristic Ramsey fringes. The full derivation is provided in the Supplementary Information.

In an ideal two-level system, the Ramsey fringe contrast decays exponentially with the homogeneous coherence time $T_2$ which is Fourier-transform limited by $T_2 = 2T_1$. In our solid-state emitter, however, slow fluctuations of the transition frequency—arising from charge noise, local electric-field drift, and strain—cause each Ramsey sequence to accumulate a slightly different phase. Averaging over many repetitions with different effective detunings leads to a Gaussian decay envelope

$$P_e(\tau) = \frac{1}{2}\left(1 + \exp\left(-\frac{\tau^2}{T_2^{*2}}\right)\cos(\omega_0\tau)\right)$$

where $T_2^*$ is the inhomogeneous coherence time. This regime is governed by

$$\frac{1}{T_2^*} = \frac{1}{T_2} + \frac{1}{T_{inh}}$$

with $T_{inh}$ quantifying additional dephasing from slow spectral wandering.

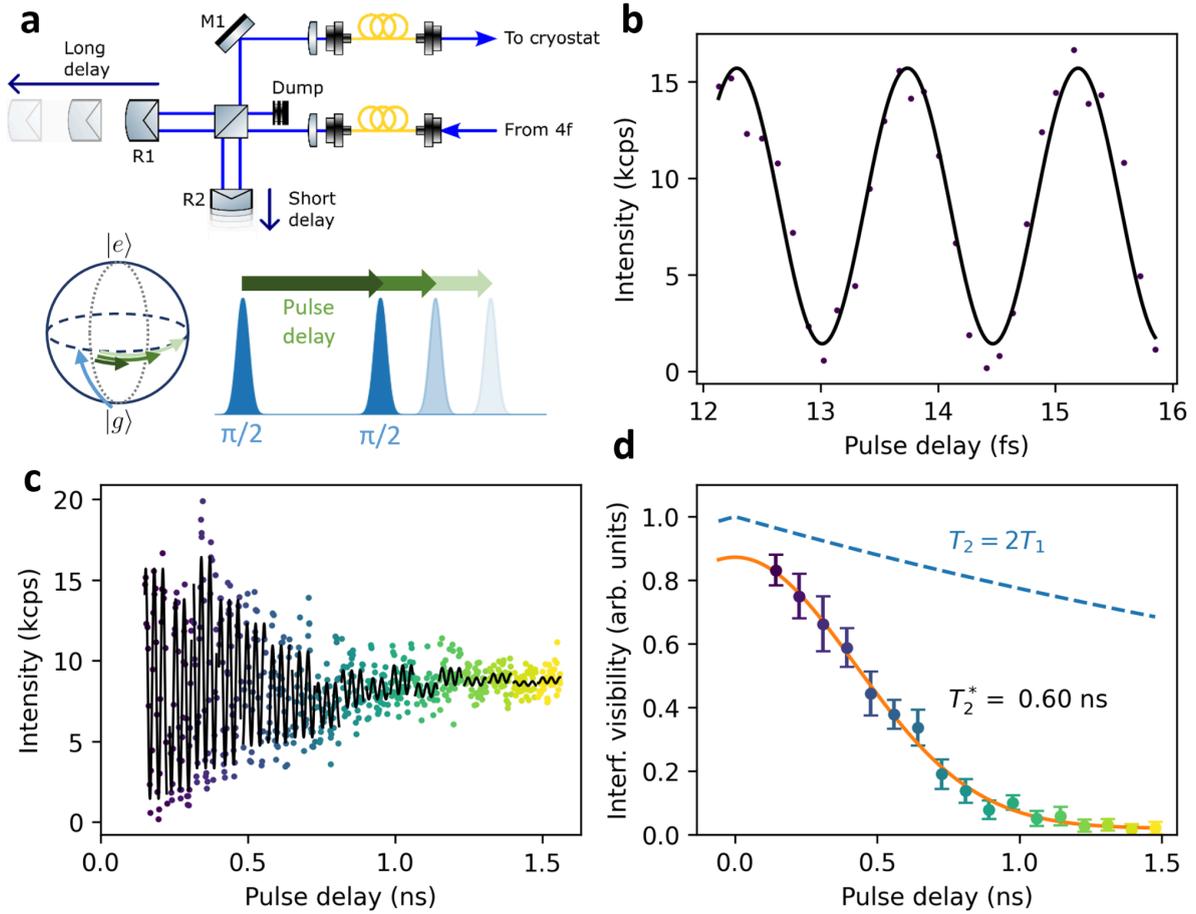

*Figure 3. Ramsey interferometry of a single B center. a) (Top) Schematic of the interferometer used to generate controlled pulse delays. A resonant π-pulse is split into two π/2 pulses at the beam splitter, and the delay between them is set by the relative position of the retroreflectors R1 and R2. (Bottom) Bloch-sphere representation of the Ramsey sequence: the first π/2 pulse prepares a coherent superposition on the equator, which then undergoes free precession during the interpulse delay. The second π/2 pulse is omitted for clarity. b) PSB fluorescence interferogram recorded with the long delay fixed at zero while scanning the short delay over several microns. A sinusoidal fit is used to extract the interference visibility. c) Interferograms measured for increasing π/2-pulse separations, implemented by incrementally increasing the long delay. For visual clarity, traces are stretched along the time axis. d) Interference visibility extracted from the interferograms in (c). A Gaussian fit yields to the inhomogeneous Ramsey coherence time $T_2^* = 0.60$ ns. For comparison, the Fourier-transform-limited envelope corresponding to the homogeneous coherence time $T_2 = 2T_1$ is shown as a blue dashed line. By definition, the inhomogeneous time $T_2^*$ is shorter than the homogeneous limit $T_2$ (see Supplementary Information).*

To implement the variable pulse delay, we route the excitation laser through a free-space Michelson interferometer, as shown in Figure 3a. The long arm of the interferometer is translated in discrete steps of 1.25 cm, while at each position the short arm is scanned over several microns using a piezo linear stage to record the interference oscillations around the chosen long-arm position. The power in each of the two pulses is verified to be 2.91 µW —corresponding to a π/2 pulse— by blocking one interferometer arm at a time and measuring the transmitted power just before the cryostat objective. Figure 3b shows the oscillating PSB fluorescence from the B center as the delay of the second π/2 pulse is incremented slightly from zero using only the short-arm scan. The high visibility of these oscillations reflects the strong quantum coherence maintained by the B center electronic state as it precesses on the Bloch sphere. For longer delays, spectral diffusion shifts the optical transition relative to the excitation laser, reducing the oscillation visibility. This behavior is shown in Figure 3c, where the horizontal axis of each interferogram has been rescaled for visual clarity.

The visibility extracted at each long-arm position is plotted in Figure 3d. The visibility decays with a Gaussian profile, characteristic of the random behaviour of spectral wandering. A Gaussian fit yields a coherence time of $T_2^* = 0.60$ ns for the B center electronic superposition state. This value is already approaching the lifetime-limited upper bound of the homogenous coherence time $2T_1 = 3.90$ ns. This is despite being measured from a B center without employing any electrical or laser-based stabilization schemes that are commonly required in other solid-state platforms[27–29]. Using the same interferometric setup, we also measure the coherence length of the resonant laser pulses to be 6.3 ps, confirming that the sustained oscillations observed over nanosecond-scale delays arise from the quantum coherence of the emitter rather than residual laser interference.

Resonant pulsed excitation enables full coherent optical control of the B center. Rabi oscillations demonstrate deterministic population transfer and strong optical isolation of the transition, while Ramsey interference directly probes the phase evolution of the optical superposition. Together, these results establish the B center as a fully addressable optical two-level system capable of controlled Bloch-sphere rotations at cryogenic temperature. Comparable benchmarks exist in other solid-state emitters—such as NV centers in diamond[30] and in SiC[31]—though these involve microwave or mechanical control of spin states. Fully optical Rabi and Ramsey sequences have been achieved in semiconductor quantum dots, where resonant pulsed driving enables deterministic rotations and phase-controlled interferometry[32]. The fact that analogous coherence can be generated here on a single B center underscores the robustness of its optical transition and highlights its relevance for coherent quantum photonics.

Ramsey interferometry quantifies the phase coherence of the optical transition by measuring the free evolution of a prepared superposition. The extracted coherence time $T_2^*$ reflects moderate spectral diffusion, consistent with slow fluctuations of the transition energy commonly observed in solid-state systems. The fact that the measured value remains within an order of magnitude of the homogeneous limit indicates that the B center coherence is intrinsically strong and only modestly perturbed by its local environment.

This work also highlights the suitability of pulsed resonant excitation as a powerful experimental approach for studying B centers. Further, coupling the emitter to optical cavities such as circular Bragg gratings or photonic crystal cavities would provide Purcell enhancement, shortening the radiative lifetime and thereby reducing the $2T_1$ coherence limit, minimizing the relative impact of spectral diffusion. Additional improvements may also be achieved through tighter spectral filtering of the excitation pulses and fine-tuning of the laser pulse bandwidth with the inhomogeneous linewidth of the B center ZPL.

Looking ahead, the coherent optical control demonstrated here suggests that B centers may integrate naturally with emerging van der Waals photonic platforms[33,34], where cavity or waveguide structures could improve photon extraction and offer Purcell enhancement. These resonators can enable practical routes for stabilizing the emission frequency and reducing the impact of slow spectral fluctuations. In this context, our measurements establish a solid foundation for assessing how controlled photonic surroundings influence the optical coherence of B centers, and they identify this defect family as a promising system for further quantum-optical experiments within two-dimensional materials.

In conclusion we demonstrate coherent optical control of a single B center in hBN, including Rabi oscillations and Ramsey interference driven directly on its optical transition. The extracted inhomogeneous coherence time of $T_2^* = 0.60$ ns indicates only moderate spectral diffusion and approaches the lifetime-limited bound, achieved without any form of active stabilization. These results establish the B center as an intrinsically coherent quantum emitter in a van der Waals material. Combined with the compatibility of hBN with photonic-cavity architectures and advanced pulsed-control schemes, this work positions the B center as a promising building block for scalable, coherent quantum photonics based on two-dimensional materials.

**Acknowledgements**

The authors acknowledge financial support from the Australian Research Council (CE200100010, FT220100053, DP250100973 ) and the Air Force Office of Scientific Research (FA2386-25-1-4044).

**References**

(1) Couteau, C.; Barz, S.; Durt, T.; Gerrits, T.; Huwer, J.; Prevedel, R.; Rarity, J.; Shields, A.; Weihs, G. Applications of Single Photons to Quantum Communication and Computing. *Nat. Rev. Phys.* **2023**, *5* (6), 326–338.

(2) Couteau, C.; Barz, S.; Durt, T.; Gerrits, T.; Huwer, J.; Prevedel, R.; Rarity, J.; Shields, A.; Weihs, G. Applications of Single Photons in Quantum Metrology, Biology and the Foundations of Quantum Physics. *Nat. Rev. Phys.* **2023**, *5* (6), 354–363.

(3) Aharonovich, I.; Englund, D.; Toth, M. Solid-State Single-Photon Emitters. *Nat. Photonics* **2016**, *10* (10), 631–641.

(4) Esmann, M.; Wein, S. C.; Antón-Solanas, C. Solid-State Single-Photon Sources: Recent Advances for Novel Quantum Materials. *Adv. Funct. Mater.* n/a (n/a), 2315936.

(5) Senellart, P.; Solomon, G.; White, A. High-Performance Semiconductor Quantum-Dot Single-Photon Sources. *Nat. Nanotechnol.* **2017**, *12* (11), 1026–1039.

(6) Müller, M.; Vural, H.; Schneider, C.; Rastelli, A.; Schmidt, O. G.; Höfling, S.; Michler, P. Quantum-Dot Single-Photon Sources for Entanglement Enhanced Interferometry. *Phys. Rev. Lett.* **2017**, *118* (25), 257402.

(7) Singh, A.; Li, Q.; Liu, S.; Yu, Y.; Lu, X.; Schneider, C.; Höfling, S.; Lawall, J.; Verma, V.; Mirin, R.; Nam, S. W.; Liu, J.; Srinivasan, K. Quantum Frequency Conversion of a Quantum Dot Single-Photon Source on a Nanophotonic Chip. *Optica* **2019**, *6* (5), 563–569.

(8) Toninelli, C.; Gerhardt, I.; Clark, A. S.; Reserbat-Plantey, A.; Götzinger, S.; Ristanović, Z.; Colautti, M.; Lombardi, P.; Major, K. D.; Deperasińska, I.; Pernice, W. H.; Koppens, F. H. L.; Kozankiewicz, B.; Gourdon, A.; Sandoghdar, V.; Orrit, M. Single Organic Molecules for Photonic Quantum Technologies. *Nat. Mater.* **2021**, *20* (12), 1615–1628.

(9) Zhang, L.; Yu, Y.-J.; Chen, L.-G.; Luo, Y.; Yang, B.; Kong, F.-F.; Chen, G.; Zhang, Y.; Zhang, Q.; Luo, Y.; Yang, J.-L.; Dong, Z.-C.; Hou, J. G. Electrically Driven Single-Photon Emission from an Isolated Single Molecule. *Nat. Commun.* **2017**, *8* (1), 580.

(10) Murtaza, G.; Colautti, M.; Hilke, M.; Lombardi, P.; Cataliotti, F. S.; Zavatta, A.; Bacco, D.; Toninelli, C. Efficient Room-Temperature Molecular Single-Photon Sources for Quantum Key Distribution. *Opt. Express* **2023**, *31* (6), 9437–9447.

(11) Schröder, T.; Trusheim, M. E.; Walsh, M.; Li, L.; Zheng, J.; Schukraft, M.; Sipahigil, A.; Evans, R. E.; Sukachev, D. D.; Nguyen, C. T.; Pacheco, J. L.; Camacho, R. M.;

Bielejec, E. S.; Lukin, M. D.; Englund, D. Scalable Focused Ion Beam Creation of Nearly Lifetime-Limited Single Quantum Emitters in Diamond Nanostructures. *Nat. Commun.* **2017**, *8* (1), 15376.

(12) Pezzagna, S.; Meijer, J. Quantum Computer Based on Color Centers in Diamond. *Appl. Phys. Rev.* **2021**, *8* (1), 011308.

(13) Wang, J.; Zhou, Y.; Wang, Z.; Rasmita, A.; Yang, J.; Li, X.; von Bardeleben, H. J.; Gao, W. Bright Room Temperature Single Photon Source at Telecom Range in Cubic Silicon Carbide. *Nat. Commun.* **2018**, *9* (1), 4106.

(14) Lukin, D. M.; Guidry, M. A.; Vučković, J. Integrated Quantum Photonics with Silicon Carbide: Challenges and Prospects. *PRX Quantum* **2020**, *1* (2), 020102.

(15) Tran, T. T.; Bray, K.; Ford, M. J.; Toth, M.; Aharonovich, I. Quantum Emission from Hexagonal Boron Nitride Monolayers. *Nat. Nanotechnol.* **2016**, *11* (1), 37–41.

(16) Sajid, A.; Ford, M. J.; Reimers, J. R. Single-Photon Emitters in Hexagonal Boron Nitride: A Review of Progress. *Rep. Prog. Phys.* **2020**, *83* (4), 044501.

(17) Shevitski, B.; Gilbert, S. M.; Chen, C. T.; Kastl, C.; Barnard, E. S.; Wong, E.; Ogletree, D. F.; Watanabe, K.; Taniguchi, T.; Zettl, A.; Aloni, S. Blue-Light-Emitting Color Centers in High-Quality Hexagonal Boron Nitride. *Phys. Rev. B* **2019**, *100* (15), 155419.

(18) Horder, J.; White, S. J. U.; Gale, A.; Li, C.; Watanabe, K.; Taniguchi, T.; Kianinia, M.; Aharonovich, I.; Toth, M. Coherence Properties of Electron-Beam-Activated Emitters in Hexagonal Boron Nitride Under Resonant Excitation. *Phys. Rev. Appl.* **2022**, *18* (6), 064021.

(19) Fournier, C.; Watanabe, K.; Taniguchi, T.; Barjon, J.; Buil, S.; Hermier, J.-P.; Delteil, A. Investigating the Fast Spectral Diffusion of a Quantum Emitter in hBN Using Resonant Excitation and Photon Correlations. *Phys. Rev. B* **2023**, *107* (19), 195304.

(20) Fournier, C.; Plaud, A.; Roux, S.; Pierret, A.; Rosticher, M.; Watanabe, K.; Taniguchi, T.; Buil, S.; Quélin, X.; Barjon, J.; Hermier, J.-P.; Delteil, A. Position-Controlled Quantum Emitters with Reproducible Emission Wavelength in Hexagonal Boron Nitride. *Nat. Commun.* **2021**, *12* (1), 3779.

(21) Gale, A.; Li, C.; Chen, Y.; Watanabe, K.; Taniguchi, T.; Aharonovich, I.; Toth, M. Site-Specific Fabrication of Blue Quantum Emitters in Hexagonal Boron Nitride. *ACS Photonics* **2022**, *9* (6), 2170–2177.

(22) Gérard, D.; Rosticher, M.; Watanabe, K.; Taniguchi, T.; Barjon, J.; Buil, S.; Hermier, J.-P.; Delteil, A. Top-down Integration of an hBN Quantum Emitter in a Monolithic Photonic Waveguide. *Appl. Phys. Lett.* **2023**, *122* (26), 264001.

(23) Nonahal, M.; Horder, J.; Gale, A.; Ding, L.; Li, C.; Hennessey, M.; Ha, S. T.; Toth, M.; Aharonovich, I. Deterministic Fabrication of a Coupled Cavity–Emitter System in Hexagonal Boron Nitride. *Nano Lett.* **2023**, *23* (14), 6645–6650.

(24) Fournier, C.; Roux, S.; Watanabe, K.; Taniguchi, T.; Buil, S.; Barjon, J.; Hermier, J.-P.; Delteil, A. Two-Photon Interference from a Quantum Emitter in Hexagonal Boron Nitride. *Phys. Rev. Appl.* **2023**, *19* (4), L041003.

(25) Hou, H.; Hua, M.; Kolluru, V. S. C.; Chen, W.-Y.; Yin, K.; Tripathi, P.; Chan, M. K. Y.; Diroll, B. T.; Gage, T. E.; Zuo, J.-M.; Wen, J. Nanometer Resolution Structure-Emission Correlation of Individual Quantum Emitters via Enhanced Cathodoluminescence in Twisted Hexagonal Boron Nitride. *Adv. Mater.* **2025**, *37* (41), e01611.

(26) Gérard, D.; Buil, S.; Hermier, J.-P.; Delteil, A. Crossover from Inhomogeneous to Homogeneous Response of a Resonantly Driven hBN Quantum Emitter. *Phys. Rev. B* **2025**, *111* (8), 085304.

(27) Prechtel, J. H.; Kuhlmann, A. V.; Houel, J.; Greuter, L.; Ludwig, A.; Reuter, D.; Wieck, A. D.; Warburton, R. J. Frequency-Stabilized Source of Single Photons from a Solid-State Qubit. *Phys. Rev. X* **2013**, *3* (4), 041006.

(28) Acosta, V. M.; Santori, C.; Faraon, A.; Huang, Z.; Fu, K.-M. C.; Stacey, A.; Simpson, D. A.; Ganesan, K.; Tomljenovic-Hanic, S.; Greentree, A. D.; Prawer, S.; Beausoleil, R. G. Dynamic Stabilization of the Optical Resonances of Single Nitrogen-Vacancy Centers in Diamond. *Phys. Rev. Lett.* **2012**, *108* (20), 206401.

(29) Hansom, J.; Schulte, C. H. H.; Matthiesen, C.; Stanley, M. J.; Atatüre, M. Frequency Stabilization of the Zero-Phonon Line of a Quantum Dot via Phonon-Assisted Active Feedback. *Appl. Phys. Lett.* **2014**, *105* (17), 172107.

(30) MacQuarrie, E. R.; Gosavi, T. A.; Moehle, A. M.; Jungwirth, N. R.; Bhave, S. A.; Fuchs, G. D. Coherent Control of a Nitrogen-Vacancy Center Spin Ensemble with a Diamond Mechanical Resonator. *Optica* **2015**, *2* (3), 233–238.

(31) Murzakhanov, F. F.; Dmitrieva, E. V.; Baibekov, E. I.; Sadovnikova, M. A.; Mamin, G. V.; Nagalyuk, S. S.; Gafurov, M. R. Mechanism of Damping of Rabi Oscillations of NV Centers in a $6H$-SiC Crystal. *Phys. Rev. B* **2025**, *111* (21), 214116.

(32) Jayakumar, H.; Predojević, A.; Huber, T.; Kauten, T.; Solomon, G. S.; Weihs, G. Deterministic Photon Pairs and Coherent Optical Control of a Single Quantum Dot. *Phys. Rev. Lett.* **2013**, *110* (13), 135505.

(33) Caldwell, J. D.; Aharonovich, I.; Cassabois, G.; Edgar, J. H.; Gil, B.; Basov, D. N. Photonics with Hexagonal Boron Nitride. *Nat. Rev. Mater.* **2019**, *4* (8), 552–567.

(34) Zotev, P. G.; Bouteyre, P.; Wang, Y.; Randerson, S. A.; Hu, X.; Sortino, L.; Wang, Y.; Shegai, T.; Gong, S.-H.; Tittl, A.; Aharonovich, I.; Tartakovskii, A. I. Nanophotonics with Multilayer van Der Waals Materials. *Nat. Photonics* **2025**, *19* (8), 788–802.